\documentclass[12pt]{article}

\usepackage[dvipdfmx]{graphicx}
\usepackage{subcaption}
\usepackage{listings,jvlisting}
\usepackage{amsmath,amssymb}
\usepackage{bm}
\usepackage{graphicx, color}
\usepackage{wrapfig}
\usepackage{cite}
\usepackage{braket}
\usepackage{hyperref}
\hypersetup{
setpagesize = false,
bookmarksnumbered = true,
bookmarksopen = true,
colorlinks = true,
linkcolor = blue,
citecolor = blue,
}
 
\usepackage{here} 
\usepackage{booktabs}
\usepackage{setspace}
\usepackage{MnSymbol} 

\title{
\begin{flushright}
\ \\*[-80pt]
\begin{minipage}{0.2\linewidth}
\normalsize
HUPD-2401 \\*[50pt]
\end{minipage}
\end{flushright}
{\Large \bf
New classification method for Equivalence Classes on $S^1/Z_2$ and $T^2/Z_3$ Orbifolds
\\*[20pt]}
}

\author{
\centerline{
Kota Takeuchi $^{1}$\footnote{k-takeuchi@hiroshima-u.ac.jp}
Tomohiro Inagaki $^{1,2,3}$\footnote{inagaki@hiroshima-u.ac.jp}}
\\*[20pt]
\centerline{
\begin{minipage}{\linewidth}
\begin{center}
$^1${\it \normalsize
Graduate School of Advanced Science and Engineering, Hiroshima University,
Higashi-Hiroshima~739-8526,~Japan \\*[5pt]
$^2${\it \normalsize
Information Media Center, Hiroshima University, Higashihiroshima 739-8521, Japan} \\
$^3${Core of Research for the Energetic Universe, Hiroshima University, Higashihiroshima 739-8526, Japan}
}
\end{center}
\end{minipage}}
\\*[50pt]}
\date{}

\begin{document}
\maketitle


\begin{abstract}
In five- and six-dimensional $U(N)$ and $SU(N)$ gauge theories compactified on $S^1/Z_2$ and $T^2/Z_3$ orbifolds, we propose a new method to classify the equivalence classes (ECs) of boundary conditions (BCs) wihtout depending on the structure of gauge transformations.
Some of the BCs are connected through gauge transformations and constitute ECs, each of which contains physically equivalent BCs.
Previous methods for classifying ECs have been used specific gauge transformations.
In this paper, we show that a geometric property of orbifolds significantly narrows down the possibilities of connecting BCs and completes the classification of ECs.
\end{abstract}


\newpage
\section{Introduction} \label{sec_intro}
The Standard Model (SM) describes the three fundamental interactions in a unified framework except for gravity and can explain many experimental results.
However, it still has several experimental and theoretical problems, such as neutrino masses, dark matter, flavor mixing, baryon asymmetry and hierarchy problems, suggesting the existence of physics beyond the Standard Model (BSM).
In recent decades, higher dimensional theories have been actively studied inspired by string theory, the highest energy physics developed in 10-dimensional space-time to unify the four fundamental interactions.
They have been proposed as one of the attractive BSM, naturally resolving the hierarchy problem and other unsolved problems\cite{ARKANIHAMED1998263, PhysRevLett.83.3370, PhysRevD.64.035002, MANTON1979141, FAIRLIE197997, DBFairlie_1979, HOSOTANI1983309, HOSOTANI1983193, HOSOTANI1989233, doi:10.1142/S021773239800276X}.

Gauge Higgs Unification (GHU) theory is one of higher dimensional theories\cite{MANTON1979141, FAIRLIE197997, DBFairlie_1979, HOSOTANI1983309, HOSOTANI1983193, HOSOTANI1989233}.
They are gauge theories in higher dimensional space-time, identifying the extra dimensional components of a gauge field with a Higgs field.
In these theories, many free parameters in SM are reduced since the Higgs potential and Yukawa interaction terms are embedded in gauge kinetic and interaction terms.
Additionally, the Higgs boson is protected by the gauge principle so that the hierarchy problem can be naturally avoided\cite{doi:10.1142/S021773239800276X}.
There are a lot of GHU models composed of various space-time structures, symmetries, field contents, and boundary conditions\cite{10.1143/PTP.103.613, 10.1143/PTP.105.999, PhysRevD.64.055003, doi:10.1142/S0217732302008988, SCRUCCA2003128, PhysRevD.69.055006, BURDMAN20033, PhysRevD.70.015010, PhysRevD.67.085012, AGASHE2005165, PhysRevD.78.096002, PhysRevD.79.079902, 10.1093/ptep/ptu146, PhysRevD.104.115018, HOSOTANI2005276, PANICO2006186, PANICO2007189, PhysRevD.98.015022, 10.1093/ptep/ptz083, PhysRevD.106.055033}.

On compactified extra dimensional space, there are numerous choices for boundary conditions (BCs) of fields.
Various BCs introduce different gauge symmetry breaking patterns and mass spectra\cite{doi:10.1142/S0217732302008988}, so that even if the same Lagrangian is given, different 4D models can be constructed depending on the choice for the BCs.
It is still unclear which BCs should be chosen without relying on phenomenological information.
This is called the arbitrariness problem of BCs\cite{doi:10.1142/5326, Quiros:2003gg, HABA2003169, 10.1143/PTP.111.265}.

This arbitrariness is partially removed by the construction of equivalence classes (ECs)\cite{HOSOTANI1989233, HABA2003169}.
In higher dimensional gauge theories, BCs are not generally gauge-invariant and some of them can be connected through gauge transformations.
Such BCs belong to the same equivalence class (EC), that is a set of BCs connected by gauge transformations.
BCs in an EC are different but produce equivalent physics.
In other words, physics is determined by ECs, not BCs.
Therefore, classifying ECs is phenomenologically and theoretically important because it contributes to more systematic understanding of models and is a crucial step towards resolving the arbitrariness problems of BCs.

Previous studies have classified ECs in 5D and 6D theories compactified on $S^1/Z_2$ and $T^2/Z_N$ $(N=2,3,4,6)$ orbifolds, which are phenomenologically appealing for naturally introducing residual gauge symmetry and 4D chiral fermions at the measurable low-energy scale\cite{HABA2003169, 10.1143/PTP.111.265, PhysRevD.69.125014, 10.1143/PTP.120.815, 10.1143/PTP.122.847, doi:10.1142/S0217751X20502061, Kawamura2023}.
However, they have found some equivalent relations between BCs by using specific gauge transformations in the extra dimensions.
We emphasize that it is not sufficient because the possibilities of new equivalent relations are left under other gauge transformations.

In this paper, we introduce a new classification method for ECs independent of the structure of gauge transformations.
We address $U(N)$ and $SU(N)$ gauge theories on $S^1/Z_2$ and $T^2/Z_3$ orbifold.
It is shown that a geometric property of the orbifolds significantly narrows down the possibilities of connecting BCs and completes the classification of ECs.
This paper is organized as follows:
In Sec.\ref{sec_property}, we define ECs on $S^1/Z_2$ and $T^2/Z_3$ orbifolds and point out the insufficiency of traditional classification methods for ECs.
In Sec.\ref{sec_S1Z2} and \ref{sec_T2Z3}, a new classification method is introduced on each orbifold.
We find that this is a general method independent of the structure of gauge transformations.
Sec.\ref{sec_concl} gives conclusions.


\section{Orbifolds and Equivalence Classes} \label{sec_property}
In this section, we review the general properties of $S^1/Z_2$ and $T^2/Z_3$ orbifolds and introduce conventional classification method for equivalence classes (ECs).

\subsection{$S^1/Z_2$ orbifold} \label{S1Z2orbi}
\begin{figure} [h]
\centering\includegraphics*{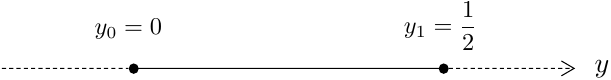}
\caption{$S^1/Z_2$ orbifold whose fundamental region is represented by the solid line with the two basic fixed points $y_0$ and $y_1$.}
\label{fig_S1Z2}
\end{figure}

Let $x^M=(x^\mu,y)$ $(\mu=0,1,2,3)$ be the five-dimensional coordinates on the Minkowski space-time $M^4$ and the orbifold $S^1/Z_2$ (Fig.\ref{fig_S1Z2}), which is identified as two points on a circle $S^1$ with radius R by parity.
The $y$-coordinate on $S^1/Z_2$ is identified as
\begin{equation} \label{S1Z2_identify}
    y \sim y+2\pi R \sim -y.
\end{equation}
Hereafter we take $2\pi R =1$ because the size of extra dimension is irrelevant to the following discussions.
The basic operators on $S^1/Z_2$ are defined by
\begin{equation} \label{Z2trans}
    \mathcal{\hat{P}}_0 : y \to -y, \quad
    \mathcal{\hat{P}}_1 : \frac{1}{2} +y \to \frac{1}{2} -y,
\end{equation}
where $\mathcal{\hat{P}}_0$ and $\mathcal{\hat{P}}_1$ are respectively parity transformations around the two fixed points,
\begin{equation} \label{S1Z2fp}
    y_0=0,\quad y_1=\frac{1}{2}.
\end{equation}
They satisfy $\mathcal{\hat{P}}^2_0 = \mathcal{\hat{P}}^2_1 = \mathcal{\hat{I}}$, where $\mathcal{\hat{I}}$ is the identity operator. 
We do not need to concern about the translation operator, $\mathcal{\hat{T}}: y \to y+1$, since it is generated by the above two operators, $\mathcal{\hat{T}}= \mathcal{\hat{P}}_1\mathcal{\hat{P}}_0$.

\vskip1.5\baselineskip
\vskip0.5\baselineskip

Let us consider $U(N)$ and $SU(N)$ gauge theories on $M^4\times S^1/Z_2$.
Although the Lagrangian must be invariant under $\mathcal{\hat{P}}_0$ and $\mathcal{\hat{P}}_1$ operations, the values of fields are not necessarily the same.
Boundary conditions (BCs) of the fields are generally non-trivial.
For example, the BCs for the four-dimensional components of 5D gauge fields, which belong to adjoint representations, are described by
(Hereafter the subscript $\mu$ on $x^\mu$ is omitted for simplicity),
\begin{gather} 
    A_{\mu}(x, -y) = P_0 A_{\mu}(x, y) P^\dag_0, \label{S1Z2_AmuBC0} \\
    A_{\mu}(x, 1-y) = P_1 A_{\mu}(x, y) P^\dag_1, \label{S1Z2_AmuBC1}
\end{gather}
where $P_i$ $(i=0,1)$ are the $N\times N$ representation matrices of the parity operators $\mathcal{\hat{P}}_i$, i.e. $P_i=P_i^{-1}=P_i^{\dag}$.
BCs on $S^1/Z_2$ are characterized by a pair of the two representation matrices $(P_0,P_1)$.

There are numerous choices for the pair of BCs satisfying unitarity and parity.
BCs produce various patterns of residual gauge symmetry and mass spectra of 4D fields\cite{doi:10.1142/S0217732302008988}, although in some cases they can yield identical physics.\cite{HABA2003169}.
BCs are generally not gauge-invariant.
For instance, the BCs for the transformed gauge fields are written as
\begin{gather} 
    A'_{\mu}(x, -y) = P'_0 A'_{\mu}(x, y) P_0^{'\dag}
    - \frac{i}{g} P'_0  \, \partial_\mu  \, P_0^{'\dag}, \label{S1Z2_A'muBC0} \\
    A'_{\mu}(x, 1-y) = P'_1 A'_{\mu}(x, y) P_1^{'\dag}
    - \frac{i}{g} P'_1  \, \partial_\mu  \, P_1^{'\dag}, \label{S1Z2_A'muBC1}
\end{gather}
where $P'_0$ and $P'_1$ are defined by
\begin{gather}  
    P'_0=\Omega(x, -y) P_0 \Omega^\dag(x, +y),\label{S1Z2_P0BC}\\
    P'_1=\Omega(x, 1-y) P_1 \Omega^\dag(x, +y),\label{S1Z2_P1BC}
\end{gather} 
where $\Omega(x,y)$ is a gauge transformation matrix.

In general, the transformed matrices $(P'_0,P'_1)$ are $y$-dependent.
If they remain constant and satisfy parity (and unitary) conditions, then they can be regarded as another pair which is transitioned from the original pair $(P_0,P_1)$.
We call this transformation \textit{BCs-connecting gauge transformation} in this paper.
The connected pairs $(P_0, P_1)$ and $(P'_0, P'_1)$ are different but yield equivalent physics.
Such BCs belong to the same equivalence class (EC), that is a set of BCs connected by BCs-connecting gauge transformations.%
\footnote{BCs in an EC are connected via the Wilson line phase, which cannot be gauged away\cite{HABA2003169}.
Physical symmetry is determined by the combination of BCs and Wilson line phases based on Hosotani mechanism\cite{HOSOTANI1983309, HOSOTANI1983193, HOSOTANI1989233}.}
Then they are represented as
\begin{equation} \label{S1Z2_ECdefinition}
    (P_0, P_1) \sim (P'_0, P'_1).
\end{equation}
The number of ECs is generally multiple for a given gauge group\cite{10.1143/PTP.111.265}, and the different ECs yield different physics.
Classifying ECs contributes to more systematic understanding of the models and is also a crucial step towards resolving the arbitrariness problems of the BCs.

The first step to classify ECs is checking simultaneous diagonalizability of basic pairs of BCs by unitary and gauge transformations.
If they are so, all ECs are characterized by sets of eigenvalues.
If they are not so, there is at least one EC composed exclusively of non-diagonal matrices pairs.
Refs.\cite{10.1143/PTP.111.265,doi:10.1142/S0217751X20502061,Kawamura2023} have proven that the representation matrices of BCs can be simultaneously diagonalized on $S^1/Z_2$, $T^2/Z_2$ and $T^2/Z_3$, but not on $T^2/Z_4$ and $T^2/Z_6$.
On $S^1/Z_2$, ECs in $U(N)$ and $SU(N)$ gauge theories are characterized by the set of $N$ eigenvalues for $P_i$, which are $\pm1$ from parity conditions:
\begin{align} \label{S1Z2_pqrs}
\begin{split}
    P_0 &= \overbrace{ (  +1, \cdots, +1, +1, \cdots, +1,-1, \cdots, -1,-1, \cdots, -1  )}^N,\\
    P_1 &= ( \underbrace{ +1, \cdots, +1 }_p , \underbrace{ -1, \cdots, -1 }_q , \underbrace{ +1, \cdots, +1 }_r , \underbrace{ -1, \cdots, -1 }_{s=N-p-q-r} ),
\end{split}
\end{align}
where $p,q,r,s$ are non-negative integers, and satisfy, $p+q+r+s=N$.
We use the notation of diagonal matrices with eigenvalues $a_1,\cdots,a_N$ as $(a_1,\cdots,a_N)$.
Each diagonal pair $(P_0,P_1)$ can be specified by $[p,q,r,s]$.

The next step is investigating the possibilities of connection between a diagonal pair and another diagonal pair, since more than one diagonal pair may exist in one EC.
On $S^1/Z_2$, the following connection between $(P_0,P_1)$ and $(P'_0,P'_1)$ has been found for $N=2$\cite{HABA2003169}:
\begin{equation} \label{S1Z2_EC_ex}
    [1,0,0,1] \sim [0,1,1,0].
\end{equation}
It is achieved by the gauge transformation with a $y$-linear parameter:
\begin{equation} \label{S1Z2_ylinear}
    \Omega(y) = \exp{\left[ i\pi y \sigma^1 \right]},
\end{equation}
where $\sigma^i$ $(i=1,2,3)$ are Pauli matrices.
However, we emphasize that this classification is not sufficient because there is a possibility of additional connections by other gauge transformations.

\subsection{$T^2/Z_3$ orbifold} \label{T2Z3orbi}
\begin{figure}[h]
\centering\includegraphics*{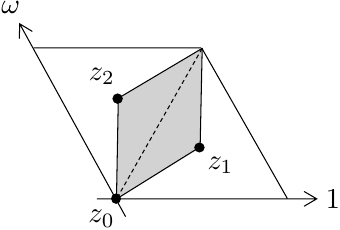}
\caption{$T^2/Z_3$ orbifold whose fundamental region is the shadow area with the three basic fixed points $z_0$, $z_1$ and $z_2$.}
\label{fig_T2Z3}
\end{figure}

Next, we consider six-dimensional space-time, $x^M=(x^\mu,y^1,y^2)$, on $M^4\times T^2/Z_3$ (Fig.\ref{fig_T2Z3}).
For 2D extra dimensions, it is convenient to introduce the complex-coordinate $z$, which is defined by, $z=y^1+iy^2$.
The 6D-coordinate is taken as $x^M=(x^\mu,z,\Bar{z})$, which $\Bar{z}$ is complex conjugate of $z$.
$T^2/Z_3$ orbifold is acquired by dividing a 2D torus $T^2$ by the $Z_3$ rotation and the $z$-coordinate satisfies the following identifications:%
\footnote{The bases on $T^2$ are arbitrary, but on $T^2/Z_3$ they are restricted to $1$ and $\omega$ because of crystallography\cite{crystal2020}.} 
\begin{equation} \label{T2Z3_identify}
    T^2:\, z \sim z+1 \sim z+\omega \quad
    Z_3:\, z \sim \omega z,
\end{equation} 
where $\omega$ is defined by $\omega=e^{2\pi i/3}$.
The three rotation operators on $T^2/Z_3$ are introduced as follows\cite{Kawamura2023}:
\begin{equation} \label{Z3trans}
    \mathcal{\hat{R}}_0 : z \to \omega z, \quad
    \mathcal{\hat{R}}_1 : z \to \omega z +1, \quad
    \mathcal{\hat{R}}_2 : z \to \omega z +1+\omega,
\end{equation}
where $\mathcal{\hat{R}}_i\,(i=0,1,2)$ represent the $Z_3$ rotations around the following fixed points:
\begin{equation} \label{T2Z3fp}
    z_0=0,\quad z_1=\frac{2+\omega}{3}, \quad z_2=\frac{1+2\omega}{3}.
\end{equation}
The rotation operators satisfy $\mathcal{\hat{R}}_i^3 =\mathcal{\hat{I}}$, where $\mathcal{\hat{I}}$ is the identity operator.
It is noted that $\mathcal{\hat{R}}_2$ is described as $\mathcal{\hat{R}}_2=\mathcal{\hat{R}}_1^2 \mathcal{\hat{R}}_0^2$.
We find that $T^2/Z_3$ orbifold has the three basic fixed points but the independent operators are the two of them, $(\mathcal{\hat{R}}_0, \mathcal{\hat{R}}_1)$.
In this paper, we do not concern about the extra translation operators, $\mathcal{\hat{T}}_i: z \to z+\omega^{i-1}\,(i=1,2,3)$, since they are also generated by,
$\mathcal{\hat{T}}_1= \mathcal{\hat{R}}_1\mathcal{\hat{R}}_0^2$ and $\mathcal{\hat{T}}_i=\mathcal{\hat{R}}_0^{i-1} \mathcal{\hat{T}}_1 \mathcal{\hat{R}}_0^{1-i}$.

\vskip\baselineskip
\vskip0.5\baselineskip

Let us look at $U(N)$ and $SU(N)$ gauge theories on $M^4\times T^2/Z_3$.
BCs for fields are characterized by a pair of representation matrices $(R_0,R_1)$ corresponding to $(\mathcal{\hat{R}}_0,\mathcal{\hat{R}}_1)$.
For instance, BCs for the four-dimensional components of 6D gauge fields are represented as
\begin{gather}
    A_{\mu}(x, \omega z, \Bar{\omega}\Bar{z}) = R_0 A_{\mu}(x, z, \Bar{z}) R^\dag_0, \label{T2Z3AmuBC0}\\
    A_{\mu}(x, \omega z+1, \Bar{\omega}\Bar{z}+1) = R_1 A_{\mu}(x, z, \Bar{z}) R^\dag_1, \label{T2Z3AmuBC1}
\end{gather}
where $R_i$ are $N\times N$ unitary matrices satisfying $Z_3$-conditions, $R_i^3=1$.
If the transformed matrices,
\begin{gather}
    R'_0=\Omega(x, \omega z, \Bar{\omega}\Bar{z}) R_0 \Omega^\dag(x, z, \Bar{z}),\\
    R'_1=\Omega(x, \omega z+1, \Bar{\omega}\Bar{z}+1) R_1 \Omega^\dag(x, z, \Bar{z}).
\end{gather}
remain constant and satisfy $Z_3$-conditions (and unitarity), then they belong to the same EC:
\begin{equation} \label{T2Z3_ECdefinition}
    (R_0, R_1) \sim (R'_0, R'_1).
\end{equation}

It has been shown in Ref.\cite{Kawamura2023} that pairs $(R_0,R_1)$ are simultaneously diagonalizable.
Therefore, ECs in $U(N)$ and $SU(N)$ gauge theories on $T^2/Z_3$ are characterized by the eigenvalues of $(R_0,R_1)$, which are $\omega$, $\omega^2$ or $1$ from $Z_3$-condition.
The diagonalized $(R_0,R_1)$ are given by
\begin{multline}
    R_0 = ( 
    \overbrace{
    \omega\,, \,\cdots\, ,\omega\,,
    \omega\,, \,\cdots\, ,\omega\,,
    \omega\,, \,\cdots\, ,\omega\,}^{n_1}\,,\,
    \overbrace{
    \omega^2, \cdots, \omega^2,
    \omega^2, \cdots, \omega^2,
    \omega^2, \cdots, \omega^2}^{n_2}\,,\, \\
    \overbrace{
    1\,,\, \cdots\,,\, 1\,\,,\,
    1\,,\, \cdots\,,\, 1\,\,,\,
    1\,,\, \cdots\,,\, 1\,}^{n_3=N-n_1-n_2})\,,
\end{multline}
\begin{multline} \label{T2Z3_nnnnnnnnn}
    R_1 = ( 
    \underbrace{\omega\,, \,\cdots\, ,\omega\,}_{n_{11}},
    \underbrace{\omega^2, \cdots, \omega^2}_{n_{12}},
    \underbrace{1\,,\, \cdots\,,\, 1\,}_{n_{13}}\,,\,
    \underbrace{\omega\,, \,\cdots\, ,\omega\,}_{n_{21}},
    \underbrace{\omega^2, \cdots, \omega^2}_{n_{22}},
    \underbrace{1\,,\, \cdots\,,\, 1\,}_{n_{23}}\,,\, \\
    \underbrace{\omega\,, \,\cdots\, ,\omega\,}_{n_{31}},
    \underbrace{\omega^2, \cdots, \omega^2}_{n_{32}},
    \underbrace{1\,,\, \cdots\,,\, 1\,}_{n_{33}}),
\end{multline}
where $n_i$ and $n_{ij}$ $(i,j=1,2,3)$ are non-negative integers, and satisfy $n_1+n_2+n_3=N$ and $n_{i1}+n_{i2}+n_{i3}=n_i$.
Each diagonal pair $(R_0,R_1)$ is specified by, $[n_{11},n_{12},n_{13}\,|\,n_{21},n_{22},n_{23}\,|\,n_{31},n_{32},n_{33}]$.
Refs.\cite{10.1143/PTP.120.815,Kawamura2023} have found the following equivalent relations:
\begin{align} \label{T2Z3_EC_ex}
    [1,0,0 \,|\, 0,1,0 \,|\, 0,0,1] 
    \sim [0,0,1 \,|\, 1,0,0 \,|\, 0,1,0] 
    \sim [0,1,0 \,|\, 0,0,1 \,|\, 1,0,0],
\end{align}
by using the specific gauge transformation,
\begin{equation} \label{T2Z3_zlinear}
    \Omega(z) = \exp{\left[i\, \frac{2\pi}{3} \left( z Y + \Bar{z} Y^\dag \right) \right]},\quad 
    Y=\begin{pmatrix} &&1\\1&&\\&1&\end{pmatrix}.
\end{equation}
However, this classification method depends on the structure of gauge transformations.
In the next section, we will show that geometric properties of orbifolds can significantly restrict the possibilities of connecting pairs of the BCs and give the complete classification for ECs on $S^1/Z_2$ and $T^2/Z_3$.

Lastly we introduce an additional BC around the other fixed point, $z_2$, because it will be used in the next section.
The gauge transformation of $R_2$ is represented as
\begin{align}
    R'_2=\Omega(x, \omega z+1+\omega, \Bar{\omega}\Bar{z}+1+\Bar{\omega}) R_2 \Omega^\dag(x, z, \Bar{z}),
\end{align}
where $R_2$ is a $N\times N$ unitary matrix with $R_2^3=1$.

\section{Classification of Equivalence Classes on $S^1/Z_2$} \label{sec_S1Z2}

\subsection{Fixed points and trace conservation on $S^1/Z_2$}
We start from the gauge transformation for a pair of the BCs on $S^1/Z_2$:
\begin{gather}  
    P'_0=\Omega(x, -y) P_0 \Omega^\dag(x, +y),\label{S1Z2_P0BC_2}\\
    P'_1=\Omega(x, 1-y) P_1 \Omega^\dag(x, +y).\label{S1Z2_P1BC_2}
\end{gather} 
It is easily checked that the traces of $P_0$ and $P_1$ are generally not conserved because the parameters of $\Omega$ and $\Omega^\dag$ are not generally consistent.
However, under BCs-connecting gauge transformations, $P'_i$ remains constant so that the right-hand side is also $y$-independent. 
It means that the trace is conserved globally as long as it holds at a particular point.
In fact, the traces of (\ref{S1Z2_P0BC_2}) and (\ref{S1Z2_P1BC_2}) are conserved at $y=y_0=0$ and $y=y_1=1/2$, respectively:
\begin{equation} \label{S1Z2_TraceCalc}
    \mathrm{tr} P'_i \,|_{y=y_i}
    = \mathrm{tr} \left[ \Omega(x, y_i) P_i \Omega^\dag(x, y_i) \right]
    = \mathrm{tr} \left[ \Omega^\dag(x, y_i) \Omega(x, y_i) P_i \right]
    = \mathrm{tr} P_i, 
\end{equation}
where we use the cyclic property of trace, $\mathrm{tr}(ABC)=\mathrm{tr}(CAB)$, and the unitarity of the transformation matrix, $\Omega^\dag(x, y_i) \Omega(x, y_i)=1$.
This trace conservation law comes from the parity invariance at each fixed point.
In the end, the traces are globally conserved under BCs-connecting gauge transformations:%
\footnote{
We note that the trace of the translation matrix $T$ on $S^1/Z_2$ is not generally conserved because it is transformed as $T' =\Omega(1+y) T \Omega^\dag(y)$.}
\begin{equation} \label{S1Z2trace}
    \mathrm{tr} P'_0 = \mathrm{tr} P_0,\quad
    \mathrm{tr} P'_1 = \mathrm{tr} P_1.
\end{equation}
In general, $U(N)$ and $SU(N)$ gauge theories on an orbifold hold the same number of the trace conservation laws as the number of the basic fixed points.

The trace conservation laws lead to the important consequence that the degeneracy of each eigenvalue for $P_i$ is invariant.
Let $n_+$ and $n_-$ be the degeneracy of eigenvalues $+1$ and $-1$ of $P_i$, and $n'_+$ and $n'_-$ denote those of eigenvalues $+1$ and $-1$ of $P'_i$.
Then the trace conservation laws (\ref{S1Z2trace}) are written as
\begin{equation} \label{S1Z2trace_eigen}
    n_+ - n_- = n'_+ - n'_-.
\end{equation}
From $n_+ + n_- = n'_+ + n'_- \,(=N)$, we obtain the invariance of the degeneracy, $n_+=n'_+$ and $n_-=n'_-$.
This means that the number of each eigenvalue $\pm1$ remains invariant under BCs-connecting gauge transformations and their permutations of the eigenvalues are only allowed.

\subsection{Equivalence Classes on $S^1/Z_2$}
Using the invariance of the degeneracy, it is possible to significantly narrow down the possibilities of connecting BCs.

Let us interchange the two eigenvalues $+1$ and $-1$ of $P_1$.
It is enough to fix $P_0$ and permute $P_1$.
For the partner $P_0$ with $(+,+)$ or $(-,-)$, we get
\begin{equation} \label{S1Z2_trivial}
    \begin{matrix}
        P_0:(\pm,\pm) \\[5pt]
        P_1:(+,-)
    \end{matrix}
    \quad \longleftrightarrow \quad
    \begin{matrix}
        P'_0:(\pm,\pm) \\[5pt]
        P'_1:(-,+).\!
    \end{matrix}
\end{equation}
This is just a trivial permutation that interchanges the bases of the eigenvalue pairs for $P_0$ and $P_1$.
The non-trivial permutation is as follows:
\begin{equation} \label{S1Z2+-+-}
    \begin{matrix}
        P_0:(+,-) \\[5pt]
        P_1:(+,-)
    \end{matrix}
    \quad \longleftrightarrow \quad
    \begin{matrix}
        P'_0:(+,-) \\[5pt]
        P'_1:(-,+).\!       
    \end{matrix}
\end{equation}
In fact, this permutation is realized by the gauge transformation (\ref{S1Z2_ylinear}).

Next, we consider the permutations of $n\,( \leq N)$ eigenvalues.
We define \textit{non-trivial n-permutation} as the permutation satisfying the following two requirements:
\begin{itemize}
  \item[(i)$\,$] All $n$ eigenvalues of $P_1$ are changed satisfying the invariance of the degeneracy.
  \item[(ii)] There is no permuted eigenvalue pair which is equal to all original pairs, i.e. $(P'_{0i},P'_{1i})\neq (P_{0j},P_{1j})$ for $i,j=1,2,\cdots n$,
\end{itemize}
where $P'_{0i},P'_{1i},P_{0i}$ and $P_{1i}$ represent the $i$-th element of $P'_{0},P'_{1},P_{0}$ and $P_{1}$, respectively.
From (i), all eigenvalues $\pm 1$ of $P_1$ change to $\mp 1$, respectively, so that the numbers of $+1$ and $-1$ of $P_1$ need to be the same.
Additionally, when $(P_{0i},P_{1i})=(+,+)$ exists, it changes to $(P'_{0i},P'_{1i})=(+,-)$, so that the other pairs of $(P_0,P_1)$ must be $(+,+)$ and $(-,-)$ from (ii).
In another case, when one pair is $(P_{0i},P_{1i})=(+,-)$, the other pairs must be $(+,-)$ and $(-,+)$.
In summary, the following non-trivial $n$-permutation is only allowed:
\begin{equation} \label{S1Z2+-_general}
    \begin{matrix}
        P_0:(+,\cdots\cdots,+\,|\,-,\cdots\cdots,-)\\[5pt]
        P_1:(+,\cdots\cdots,+\,|\,-,\cdots\cdots,-)
    \end{matrix}
    \quad \longleftrightarrow \quad
    \begin{matrix}
        P_0:(+,\cdots\cdots,+\,|\,-,\cdots\cdots,-)\\[5pt]
        P_1:(-,\cdots\cdots,-\,|\,+,\cdots\cdots,+),\!
    \end{matrix}
\end{equation}
where the numbers of $+1$ and $-1$ are the same.
We find that this is just a repetition of the 2-permutation (\ref{S1Z2+-+-}), and can be realized by using the gauge transformation (\ref{S1Z2_ylinear}).

From the above discussion on $S^1/Z_2$ orbifold, it is concluded that the possible equivalent relation is only
\begin{align} \label{S1Z2_ECrelation}
\begin{split}
    \left[ \, p,q,r,s \, \right] 
    &\sim \left[ \, p-1,q+1,r+1,s-1 \, \right]\quad \text{for}\,\, p,s\geq1,\\
    &\sim \left[ \, p+1,q-1,r-1,s+1 \, \right]\quad \text{for}\,\, q,r\geq1,
\end{split}
\end{align}
using the notation introduced in (\ref{S1Z2_pqrs}).
Therefore, we obtain a sufficient classification independent of the structure of gauge transformations.


\section{Classification of Equivalence Classes on $T^2/Z_3$} \label{sec_T2Z3}
On $T^2/Z_3$, traces of the representation matrices $R_0$ and $R_1$, associated with the fixed points $z_0$ and $z_1$, are conserved as in the case of $S^1/Z_2$.
Furthermore, $T^2/Z_3$ has the other basic fixed point $z_2$, which produce another condition.

\subsection{Fixed points and trace conservation on $T^2/Z_3$}
Let us consider gauge transformations of a pair $(R_0,R_1)$ on $T^2/Z_3$:
\begin{align}
    R'_0 &=\Omega(\omega z) R_0 \Omega^\dag(z) \label{transR0},\\
    R'_1 &=\Omega(\omega z +1) R_1 \Omega^\dag(z).
    \label{transR1}
\end{align}
The traces of (\ref{transR0}) and (\ref{transR1}) are conserved at the fixed points $z=z_0=0$ and $z=z_1=(2+\omega)/3$, respectively.
It comes from the rotational invariance at each fixed point.
From the same calculation as (\ref{S1Z2_TraceCalc}) in $S^1/Z_2$, the traces are globally conserved under BCs-connecting gauge transformations:%
\footnote{The traces of the translation matrices $T_1,T_2,T_3$ on $T^2/Z_3$ are not generally conserved, as $T$ on $S^1/Z_2$.}
\begin{equation} \label{T2Z3trace}
    \mathrm{tr} R'_0 = \mathrm{tr} R_0,\quad
    \mathrm{tr} R'_1 = \mathrm{tr} R_1.
\end{equation}

There is the other basic fixed point, $z=z_2=(1+2\omega)/3$ on $T^2/Z_3$ (See Fig.\ref{fig_T2Z3}).
The rotation matrix around $z_2$ is represented by $R_2$, which is transformed as
\begin{equation} \label{transR2}
    R'_2= \Omega(\omega z +1 +\omega) R_2 \Omega^\dag(z). 
\end{equation}
The trace of $R_2$ is conserved at $z=z_2$, so that it is globally conserved under BCs-connecting gauge transformations.
The rotation matrix $R_2$ is described by $R_0$ and $R_1$, $R_2=(R_0 R_1)^\dag$ in $U(N)$ and $SU(N)$ gauge theories.
Therefore, the trace conservation law of $R_2$ yields
\begin{equation} \label{T2Z3traceR2}
    \mathrm{tr} R'_0 R'_1 = \mathrm{tr} R_0 R_1.
\end{equation}

Due to the trace conservation laws of $R_i$, the degeneracy of each eigenvalue is preserved.%
\footnote{In the case of $T^2/Z_4$ and $T^2/Z_6$, the traces of rotation matrices are conserved, but can not straightforwardly lead to the invariance of the degeneracy of the eigenvalues because they include $Z_2$ or $Z_3$ symmetry.}
The number of eigenvalues, $\omega,\omega^2,\omega^3(=1)$ is respectively denoted as $n_1,n_2,n_3$.
Those of transformed matrices are written as $n'_1,n'_2,n'_3$.
From the trace conservation law, we get
\begin{equation} \label{T2Z3trace_eigen}
    \quad (n_1-n'_1)\omega + (n_2-n'_2)\omega^2 + (n_3-n'_3) =0.
\end{equation}
With $n_1+n_2+n_3=n'_1+n'_2+n'_3\,(=N)$ Eq.(\ref{T2Z3trace_eigen}) requires $\Delta n = (n_1-n'_1) = (n_2-n'_2) = (n_3-n'_3)$=0, so that we obtain $n_1=n'_1$, $n_2=n'_2$, $n_3=n'_3$.
Similarly, Eq.(\ref{T2Z3traceR2}) implies that the degeneracy of eigenvalues for the product $R_0R_1$ is invariant.

Therefore, in classifying ECs on $T^2/Z_3$, we investigate the possibilities of permutations of pairs $(R_0,R_1)$ with the additional condition of the product $R_0R_1$.

\subsection{Equivalence Classes on $T^2/Z_3$}
Let us consider the permutation of eigenvalues for $R_0$, $R_1$, and $R_0R_1$.
It is enough to fix $R_0$ and permute $R_1$.
First, we discuss the interchange of two eigenvalues: 
\begin{equation} \label{T2Z3ww}
    \begin{matrix}
        R_0:(\omega^i,\omega^j) \\[5pt]
        R_1:(\omega^k,\omega^l)
    \end{matrix}
    \quad \longleftrightarrow \quad
    \begin{matrix}
        R'_0:(\omega^i,\omega^j) \\[5pt]
        R'_1:(\omega^l,\omega^k),\!\!
    \end{matrix}
\end{equation}
where $i,j,k,l\in \{1,2,3\}$.
It is restricted to $i=j$ or $k=l$ due to the trace conservation law of the product $R_0R_1$.
In this case, (\ref{T2Z3ww}) is a trivial permutation that interchanges the bases of the eigenvalue pairs for $R_0$ and $R_1$.
Therefore, we conclude that non-trivial 2-permutation cannot be realized on $T^2/Z_3$.

Next, let us consider 3-permutation.
For $R_0=(\omega^i,\omega^i,\omega^i)$, any permutation becomes clearly trivial.
If $R_0$ has two kinds of the eigenvalues, $\omega^i,\omega^j$ $(\omega^i\neq\omega^j)$, we can perform
\begin{equation} \label{T2Z3www_trivial}
    \begin{matrix}
        R_0:(\omega^i,\omega^i,\omega^j) \\[5pt]
        R_1:(\omega^k,\omega^l,\omega^m) 
    \end{matrix}
    \quad \longleftrightarrow \quad
    \begin{matrix}
        R_0:(\omega^i,\omega^i,\omega^j) \\[5pt]
        R_1:(\omega^m,\omega^k,\omega^l).
    \end{matrix}
\end{equation}
This is not non-trivial 3-permutation because $(R_0,R_1)=(\omega^i,\omega^k)$ remains unchanged.
In general, both $R_0$ and $R_1$ need to possess the three kinds of eigenvalues $\omega$, $\omega^2$ and $1$, to achieve non-trivial $n$-permutation (See Appendix \ref{App_T2Z3}).
There are two possibilities:
\begin{equation} \label{T2Z3www_normal}
    \text{normal:}\quad\,
    \begin{matrix}
        R_0:(\omega,\omega^2,1) \\[5pt]
        R_1:(\omega,\omega^2,1)
    \end{matrix}
    \quad \longleftrightarrow \quad
    \begin{matrix}
        R_0:(\omega,\omega^2,1) \\[5pt]
        R_1:(1,\omega,\omega^2)
    \end{matrix}
    \quad \longleftrightarrow \quad
    \begin{matrix}
        R_0:(\omega,\omega^2,1) \\[5pt]
        R_1:(\omega^2,1,\omega),\!
    \end{matrix}
\end{equation}
\begin{equation} \label{T2Z3www_inverted}
    \text{inverted:} \quad\!\!
    \begin{matrix}
        R_0:(\omega,\omega^2,1) \\[5pt]
        R_1:(\omega,1,\omega^2)
    \end{matrix}
    \quad \longleftrightarrow \quad
    \begin{matrix}
        R_0:(\omega,\omega^2,1) \\[5pt]
        R_1:(\omega^2,\omega,1)
    \end{matrix}
    \quad \longleftrightarrow \quad
    \begin{matrix}
        R_0:(\omega,\omega^2,1) \\[5pt]
        R_1:(1,\omega^2,\omega).\!
    \end{matrix}
\end{equation}
The normal type (\ref{T2Z3www_normal}) conserves the trace of $R_0R_1$, but the inverted type (\ref{T2Z3www_inverted}) does not.
Therefore, the normal type is only allowed, which is achieved by the gauge transformation (\ref{T2Z3_zlinear}).

In Appendix \ref{App_T2Z3}, we show that the non-trivial $n\, (\leq N)$-permutation on $T^2/Z_3$ is just a repetition of the normal type permutation (\ref{T2Z3www_normal}).
Finally, without depending on the structure of gauge transformations, it is proven that the possible equivalent relations on $T^2/Z_3$ are written as
\begin{align} \label{T2Z3_ECrelation}
\begin{split}
    &\left[\, 
    n_{11},n_{12},n_{13}\,|\, 
    n_{21},n_{22},n_{23}\,|\,
    n_{31},n_{32},n_{33}\, \right] \\
    &\sim \left[\, 
    n_{11}-1,n_{12}+1,n_{13}\,|\, 
    n_{21},n_{22}-1,n_{23}+1\,|\,
    n_{31}+1,n_{32},n_{33}-1\, \right]
    \,\, \text{for}\,\, n_{11},n_{22},n_{33},\geq1,\\
    &\sim \left[\, 
    n_{11}+1,n_{12}-1,n_{13}\,|\, 
    n_{21},n_{22}+1,n_{23}-1\,|\,
    n_{31}-1,n_{32},n_{33}+1\, \right]
    \,\, \text{for}\,\, n_{12},n_{23},n_{31},\geq1,\\
    &\sim \left[\, 
    n_{11}-1,n_{12},n_{13}+1\,|\, 
    n_{21}+1,n_{22}-1,n_{23}\,|\,
    n_{31},n_{32}+1,n_{33}-1\, \right]
    \,\, \text{for}\,\, n_{11},n_{22},n_{33},\geq1,\\
    &\sim \left[\, 
    n_{11}+1,n_{12},n_{13}-1\,|\, 
    n_{21}-1,n_{22}+1,n_{23}\,|\,
    n_{31},n_{32}-1,n_{33}+1\, \right]
    \,\, \text{for}\,\, n_{13},n_{21},n_{32},\geq1,\\
    &\sim \left[\, 
    n_{11},n_{12}-1,n_{13}+1\,|\, 
    n_{21}+1,n_{22},n_{23}-1\,|\,
    n_{31}-1,n_{32}+1,n_{33}\, \right]
    \,\, \text{for}\,\, n_{12},n_{23},n_{31},\geq1,\\
    &\sim \left[\, 
    n_{11},n_{12}+1,n_{13}-1\,|\, 
    n_{21}-1,n_{22},n_{23}+1\,|\,
    n_{31}+1,n_{32}-1,n_{33}\, \right]
    \,\, \text{for}\,\, n_{13},n_{21},n_{32},\geq1,
\end{split}
\end{align}
using the notation introduced in (\ref{T2Z3_nnnnnnnnn}).


\section{Conclusion} \label{sec_concl}
We have studied equivalence classes (ECs) of boundary conditions (BCs) on $S^1/Z_2$ and $T^2/Z_3$ orbifolds in $U(N)$ and $SU(N)$ gauge theories.
It has been shown that the traces of the representation matrices for BCs are conserved under BCs-connecting gauge transformations.
It is interpreted as geometric properties of the orbifolds because it comes from the rotational (parity) invariance at each fixed point.
These trace conservation laws strongly narrow down the possibilities of connecting BCs, and only allow the permutation (\ref{S1Z2+-+-}) on $S^1/Z_2$ and (\ref{T2Z3www_normal}) on $T^2/Z_3$ and their repetitions.
As a result, we have completed classifying their ECs without depending on the structure of gauge transformations.
Our proof shows that the previous classification method using specific gauge transformations is sufficient on $S^1/Z_2$ and $T^2/Z_3$.

Our classification method is expected to have a wide range of applications.
First, it will be applied to the other 6D orbifolds, $T^2/Z_2$, $T^2/Z_4$, $T^2/Z_6$, which have the same number of trace conservation laws as the number of basic fixed points.
Yet it is a bit challenging because, on $T^2/Z_4$ and $T^2/Z_6$, the trace conservation laws can change the degeneracy of each eigenvalue.
Also there are off-diagonal ECs due to $Z_2$ or $Z_3$ sub-symmetry\cite{Kawamura2023}.
On the other hand, on $T^2/Z_2$, while the trace conservation laws lead to the invariance of its degeneracy as in the case of $S^1/Z_2$ and $T^2/Z_3$, there are not two but three basic operators\cite{Kawamura2023}.
We need to deal with a pair of three representation matrices $(P_0,P_1,P_2)$ on $T^2/Z_2$.

We can also apply our classification method to various theories with other gauge groups.
The trace conservation laws are satisfied in various gauge theories.
For instance, in $O(N)$ and $SO(N)$ gauge theories, a rotation matrix is transformed as
\begin{equation}
    R'_0=\Omega(x, \omega z, \Bar{\omega}\Bar{z}) R_0 \Omega^\mathrm{T}(x, z, \Bar{z}),
\end{equation}
where $\mathrm{T}$ denotes the transpose.
We can see that the trace of $R_0$ is also conserved under BCs-connecting gauge transformations.
Such a constraint may be able to classify ECs.
We hope to report a general classification for the other gauge groups and orbifolds in the future. 


\appendix
\section{Permutations of $n$ eigenvalues on $T^2/Z_3$} \label{App_T2Z3}
In this section, we investigate how the three trace conservation laws of $R_0$, $R_1$ and $R_0R_1$ restrict permutations of $n\,(\leq N)$ eigenvalues.
We permute $R_1$ with $R_0$ fixed without loss of generality.
Then \textit{non-trivial n-permutation} on $T^2/Z_3$ is defined by the permutation satisfying the following two requirements:
\begin{itemize}
      \item[(i)$\,\,$] All $n$ eigenvalues of $R_1$ are changed satisfying the invariance of the degeneracy.
  \item[(ii)$\,$] There is no permuted eigenvalue pair which is equal to all original pairs, i.e. $(R'_{0i},R'_{1i})\neq (R_{0j},R_{1j})$ for $i,j=1,2,\cdots n$.
  \item[(iii)] The degeneracy of the product $R_0R_1$ and $R'_0R'_1$ remains the same.
\end{itemize}
$R'_{0i},R'_{1i},R_{0i}$ and $R_{1i}$ represent the $i$-th element of $R'_{0},R'_{1},R_{0}$ and $R_{1}$, respectively.

\subsubsection*{Both $R_0$ and $R_1$ must possess all kinds of eigenvalues, $\omega$, $\omega^2$ and $1$.}
First, if $R_0$ has only one type of eigenvalue, any permutation of $R_1$ eigenvalues is the trivial transformation that interchanges the bases of the eigenvalue pairs for $R_0$ and $R_1$.

Second, we consider the case that $R_0$ has the two kinds of eigenvalues, $\omega^i, \omega^j$ ($i\neq j$ (mod $3$)).
As the permutations satisfying (i) and (ii), there are two possibilities for $a_1 \neq a_2\neq b_1\neq a_1$ (mod $3$):
\begin{align}
    R_1&=(\omega^{a_1},\cdots, \omega^{a_1}, \omega^{a_2}, \cdots, \omega^{a_2} \,|\, \omega^{b_1},\cdots, \omega^{b_1}),
    \label{2kind_1} \\
    R_1&=(\omega^{a_1},\cdots, \omega^{a_1} \,|\, \omega^{b_1},\cdots, \omega^{b_1}),
    \label{2kind_2}
\end{align}
where the left and right blocks are paired with $\omega^i$ and $\omega^j$ of $R_0$, respectively.
We exclude the cases with all three kinds of eigenvalues in the left or right block, since this violates (ii).
(\ref{2kind_1}) and (\ref{2kind_2}) should be permuted to
\begin{align}
    R'_1&=(\omega^{b_1},\cdots, \omega^{b_1} \,|\, \omega^{a_1},\cdots, \omega^{a_1}, \omega^{a_2}, \cdots, \omega^{a_2}), \\
    R'_1&=(\omega^{b_1},\cdots, \omega^{b_1} \,|\, \omega^{a_1},\cdots, \omega^{a_1}).
\end{align}
It means that the numbers of eigenvalues in each block are required to be equal.
Both cases do not satisfy (iii) because $R_0R_1$ has $\omega^{j+b_1}$, but $R'_0R'_1$ does not.
Therefore, $R_0$ needs to possess all kinds of eigenvalues to realize non-trivial $n$-permutation.
We find that $R_1$ must also possess $\omega$, $\omega^2$ and $1$ if we interchange the roles of $R_0$ and $R_1$.

From the above discussion, it is enough to consider the following four types of sets of eigenvalues:
\begin{align}
    \text{(A)}\,\,
    R_1&=(\omega^{a_1},\cdots, \omega^{a_1} \,|\, \omega^{b_1},\cdots, \omega^{b_1} \,|\, \omega^{c_1}, \cdots, \omega^{c_1}), \\
    \text{(B)}\,\,
    R_1&=(\omega^{a_1},\cdots, \omega^{a_1}, \omega^{a_2},\cdots, \omega^{a_2} \,|\, \omega^{b_1},\cdots, \omega^{b_1} \,|\, \omega^{c_1}, \cdots, \omega^{c_1}), \\
    \text{(C)}\,\,
    R_1&=(\omega^{a_1},\cdots, \omega^{a_1}, \omega^{a_2},\cdots, \omega^{a_2} \,|\, \omega^{b_1},\cdots, \omega^{b_1}, \omega^{b_2},\cdots, \omega^{b_2} \,|\, \omega^{c_1}, \cdots, \omega^{c_1}), \\
    \text{(D)}\,\,
    R_1&=(\omega^{a_1},\cdots, \omega^{a_1}, \omega^{a_2},\cdots, \omega^{a_2} \,|\, \omega^{b_1},\cdots, \omega^{b_1}, \omega^{b_2},\cdots, \omega^{b_2} \,|\, \omega^{c_1}, \cdots, \omega^{c_1}, \omega^{c_2}, \cdots, \omega^{c_2}),
\end{align}
where $a_1 \neq a_2$, $b_1 \neq b_2$, $c_1 \neq c_2$ (mod $3$), and the left, middle and right blocks are respectively paired with $\omega$, $\omega^2$ and $1$ in $R_0$.
Let $n_1$, $n_2$ and $n_3$ be the number of eigenvalues in left, middle and right blocks, respectively.
They satisfy $n_1>0$, $n_2>0$, $n_3>0$ and $n_1+n_2+n_3=n$. 

\subsubsection*{Type (B) permutations violate the trace conservation of $R_0R_1$.}
Type (B) is divided into the cases of $c_1=b_1$ and $c_1\neq b_1$.
In the case of $c_1=b_1$, the permutation satisfying (i) and (ii) is
\begin{align}
    R_1&=(\omega^{a_1},\cdots, \omega^{a_1}, \omega^{a_2},\cdots, \omega^{a_2} \,|\, \omega^{b_1},\cdots, \omega^{b_1} \,|\, \omega^{b_1}, \cdots, \omega^{b_1}) \notag \\
    \to \,\,
    R'_1&=(\omega^{b_1},\cdots, \omega^{b_1} \,|\, \omega^{a_1},\cdots, \omega^{a_1}, \omega^{a_2}, \cdots, \omega^{a_2} \,|\, \omega^{a_1},\cdots, \omega^{a_1}, \omega^{a_2}, \cdots, \omega^{a_2}),
\end{align}
where $n_i$ $(i=1,2,3)$ satisfy $n_1=n_2 + n_3$.
However, it violates (iii).
Actually $R_0R_1$ does not have $\omega^{1+b_1}$ that is one of the eigenvalues for $R'_0R'_1$, because of $a_1\neq b_1$ and $a_2\neq b_1$.

In the case of $c_1\neq b_1$, $c_1$ must be equal to $a_1$ or $a_2$.
We take $c_1=a_1$ without loss of generality.
From (i) and (ii), $R_1$ is permuted to
\begin{align}
    R_1&=(\omega^{a_1},\cdots, \omega^{a_1}, \omega^{a_2},\cdots, \omega^{a_2} \,|\, \omega^{b_1},\cdots, \omega^{b_1} \,|\, \omega^{a_1}, \cdots, \omega^{a_1}) \notag \\
    \to \,\,
    R'_1&=(\omega^{b_1},\cdots, \omega^{b_1} \,|\, \omega^{a_1},\cdots, \omega^{a_1}, \omega^{a_2}, \cdots, \omega^{a_2} \,|\, \omega^{a_2}, \cdots, \omega^{a_2}),
\end{align}
where $n_i$ $(i=1,2,3)$ must satisfy $n_1=n_2 > n_3$.
To satisfy (iii), all $\omega^{2+b_1}$ of $R_0R_1$ must be consistent with $\omega^{0+a_2}$ of $R'_0R'_1$, so that $n_2 \leq n_3$ is required.
Therefore, non-trivial $n$-permutation cannot be realized in Type (B).

\subsubsection*{Type (C) permutations violate the trace conservation of $R_0R_1$.}
Type (C) is divided into the case that $c_1$ is equal to one of $(a_1,a_2,b_1,b_2)$ and the case that $c_1$ is not equal to all.

Let us consider the former case.
We set $c_1=a_1$ and $a_2=b_2$ without loss of generality.
$R_1$ is permuted to
\begin{align}
    R_1&=(\omega^{a_1},\cdots, \omega^{a_1}, \omega^{a_2},\cdots, \omega^{a_2} \,|\, \omega^{b_1},\cdots, \omega^{b_1}, \omega^{a_2},\cdots, \omega^{a_2} \,|\, \omega^{a_1}, \cdots, \omega^{a_1}) \notag \\
    \to \,\,
    R'_1&=(\omega^{b_1},\cdots, \omega^{b_1} \,|\, \omega^{a_1},\cdots, \omega^{a_1} \,|\, \omega^{a_2},\cdots, \omega^{a_2}, \omega^{b_1}, \cdots, \omega^{b_1}).
\end{align}
From (i) and (ii), all eigenvalues in the left block of $R'_1$ are $\omega^{b_1}$, so that $n_1<n_2$ is required.
However, $n_2<n_1$ is also required because all $\omega^{2+a_1}$ of $R'_0R'_1$ must be consistent with $\omega^{1+a_2}$ of $R_0R_1$ from (iii).
These are contradicted.

In the latter case, we take $a_1=b_1$ and $a_2=b_2$ without loss of generality.
From (i) and (ii), $R_1$ is permuted to
\begin{align}
    R_1&=(\omega^{a_1},\cdots, \omega^{a_1}, \omega^{a_2},\cdots, \omega^{a_2} \,|\, \omega^{a_1},\cdots, \omega^{a_1}, \omega^{a_2},\cdots, \omega^{a_2} \,|\, \omega^{c_1}, \cdots, \omega^{c_1}) \notag \\
    \to \,\,
    R'_1&=(\omega^{c_1},\cdots, \omega^{c_1} \,|\, \omega^{c_1},\cdots, \omega^{c_1} \,|\, \omega^{a_1},\cdots, \omega^{a_1}, \omega^{a_2}, \cdots, \omega^{a_2}),
\end{align}
where $n_i$ $(i=1,2,3)$ satisfy $n_1+n_2=n_3$.
$R_0R_1$ has $\omega^{0+c_1}$, but $R'_0R'_1$ does not because of $c_1\neq a_1$ and $c_1\neq a_2$, so that (iii) is not satisfied.
Therefore, we conclude that non-trivial $n$-permutation cannot be realized for Type (C).

\subsubsection*{Type (D) can be permuted to Type (A).}
In Type (D), there are two possibilities satisfying (i) and (ii):
\begin{align} 
    R_1&=(\omega^{a+1},\cdots, \omega^{a+1}, \omega^{a-1},\cdots, \omega^{a-1} \,|\, \omega^{a+1},\cdots, \omega^{a+1}, \omega^{a},\cdots, \omega^{a} \,|\, \omega^{a},\cdots, \omega^{a}, \omega^{a-1},\cdots, \omega^{a-1}),
    \label{typeD1} \\
    R_1&=(\omega^{a+1},\cdots, \omega^{a+1}, \omega^{a-1},\cdots, \omega^{a-1} \,|\, \omega^{a-1},\cdots, \omega^{a-1}, \omega^{a},\cdots, \omega^{a} \,|\, \omega^{a},\cdots, \omega^{a}, \omega^{a+1},\cdots, \omega^{a+1}).
    \label{typeD2}
\end{align}
The upper case (\ref{typeD1}) is permuted to
\begin{align}
    R'_1&=(\omega^{a},\cdots, \omega^{a} \,|\, \omega^{a-1},\cdots, \omega^{a-1} \,|\, \omega^{a+1}, \cdots, \omega^{a+1}).
\end{align}
However, it does not satisfy (iii) because $R_0R_1$ does not have $\omega^{1+a}$ of $R'_0R'_1$.

The lower case (\ref{typeD2}) realizes the non-trivial permutation:
\begin{align}
    R_1&=(
    \overbrace{\omega^{a+1},\cdots, \omega^{a+1}}^{m_1},
    \overbrace{\omega^{a-1},\cdots, \omega^{a-1}}^{m'_1} 
    \,|\, 
    \overbrace{\omega^{a-1},\cdots, \omega^{a-1}}^{m_2} ,
    \overbrace{\omega^{a},\cdots, \omega^{a}}^{m'_2} 
    \,|\, 
    \overbrace{\omega^{a},\cdots, \omega^{a}}^{m_3}, 
    \overbrace{\omega^{a+1},\cdots, \omega^{a+1}}^{m'_3}) \notag \\
    \to \,\,
    R'_1&=(\omega^{a},\cdots, \omega^{a} \,|\, \omega^{a+1},\cdots, \omega^{a+1} \,|\, \omega^{a-1}, \cdots, \omega^{a-1}),
    \label{typeD3}
\end{align}
where $m_i$ and $m'_i$ $(i=1,2,3)$ respectively denote the number of $\omega^{a+i}$ and $\omega^{a+i+1}$ of $R_1$ paired with $\omega^{i}$ of $R_0$, and satisfy $m_i+m'_i=n_i$.
To check (i) (ii) and (iii) conditions, we consider the permutation of $R_0R_1$:
\begin{align}
    R_0R_1&=(
    \overbrace{\omega^{a-1},\cdots, \omega^{a-1}}^{m_1},
    \overbrace{\omega^{a},\cdots, \omega^{a}}^{m'_1} 
    \,|\, 
    \overbrace{\omega^{a+1},\cdots, \omega^{a+1}}^{m_2} ,
    \overbrace{\omega^{a-1},\cdots, \omega^{a-1}}^{m'_2} 
    \,|\, 
    \overbrace{\omega^{a},\cdots, \omega^{a}}^{m_3}, 
    \overbrace{\omega^{a+1},\cdots, \omega^{a+1}}^{m'_3}) \notag \\
    \to \,\,
    R'_0R'_1&=(\omega^{a+1},\cdots, \omega^{a+1} \,|\, \omega^{a},\cdots, \omega^{a} \,|\, \omega^{a-1}, \cdots, \omega^{a-1}),
\end{align}
The trace conservation laws of $R_0$, $R_1$ and $R_0R_1$ lead to
\begin{align}
    n_1 = m_1 + m'_1 = m_3 + m'_2 = m_2 + m'_3, \label{n1} \\
    n_2 = m_2 + m'_2 = m_1 + m'_3 = m_2 + m'_1, \label{n2} \\
    n_3 = m_3 + m'_3 = m_2 + m'_1 = m_1 + m'_2. \label{n3}
\end{align}
From (\ref{n1}), (\ref{n2}) and (\ref{n3}), we obtain
\begin{equation} \label{typeDcond}
    m_1=m_2=m_3, \quad m'_1=m'_2=m'_3, \quad n_1=n_2=n_3=n/3.
\end{equation}
The non-trivial $n$-permutation (\ref{typeD3}) with the conditions (\ref{typeDcond}) is only possible.
We find that this is the permutation between Type (D) and Type (A).

\subsubsection*{Type (A) can be permuted to Type (D) or itself.}
It is obvious that Type (A) can be permuted to Type (D) from the previous discussion.
Type (A) can not only move to Type (D), but also to Type (A) itself.

From (i) and (ii), there are two possibilities of the permutations closed in Type (A):
\begin{align} \label{A_upper}
    R_1&=(\omega^{a},\cdots, \omega^{a} \,|\, \omega^{a+1},\cdots, \omega^{a+1} \,|\, \omega^{a-1}, \cdots, \omega^{a-1}) \notag \\
    \to \,\,
    R'_1&=(\omega^{a-1},\cdots, \omega^{a-1} \,|\, \omega^{a},\cdots, \omega^{a} \,|\, \omega^{a+1}, \cdots, \omega^{a+1}) \\
    \to \,\,
    R''_1&=(\omega^{a+1},\cdots, \omega^{a+1} \,|\, \omega^{a-1},\cdots, \omega^{a-1} \,|\, \omega^{a}, \cdots, \omega^{a}), \notag
\end{align}
\begin{align} \label{A_lower}
    R_1&=(\omega^{a},\cdots, \omega^{a} \,|\, \omega^{a-1},\cdots, \omega^{a-1} \,|\, \omega^{a+1}, \cdots, \omega^{a+1}) \notag \\
    \to \,\,
    R'_1&=(\omega^{a+1},\cdots, \omega^{a+1} \,|\, \omega^{a},\cdots, \omega^{a} \,|\, \omega^{a-1}, \cdots, \omega^{a-1}) \\
    \to \,\,
    R''_1&=(\omega^{a-1},\cdots, \omega^{a-1} \,|\, \omega^{a+1},\cdots, \omega^{a+1} \,|\, \omega^{a}, \cdots, \omega^{a}), \notag
\end{align}
where $n_1=n_2=n_3$. 
The upper case (\ref{A_upper}) satisfies (iii), but the lower case (\ref{A_lower}) does not, so that (\ref{A_upper}) is only allowed.

Finally, there are two patterns of non-trivial $n$-permutations on $T^2/Z_3$:
\begin{align} 
    R_1&=(
    \overbrace{\omega^{a},\cdots, \omega^{a}}^{n_1} \,|\, 
    \overbrace{\omega^{a+1},\cdots, \omega^{a+1}}^{n_2} \,|\, 
    \overbrace{\omega^{a-1}, \cdots, \omega^{a-1}}^{n_3} 
    ) \\
    \leftrightarrow \,\,
    R'_1&=(
    \underbrace{\omega^{a+1},\cdots, \omega^{a+1}}_{m_1},
    \underbrace{\omega^{a-1},\cdots, \omega^{a-1}}_{m'_1} 
    \,|\, 
    \underbrace{\omega^{a-1},\cdots, \omega^{a-1}}_{m_2} ,
    \underbrace{\omega^{a},\cdots, \omega^{a}}_{m'_2} 
    \,|\, 
    \underbrace{\omega^{a},\cdots, \omega^{a}}_{m_3}, 
    \underbrace{\omega^{a+1},\cdots, \omega^{a+1}}_{m'_3}
    ), \notag\\
    R_1&=(\omega^{a},\cdots, \omega^{a} \,|\, \omega^{a+1},\cdots, \omega^{a+1} \,|\, \omega^{a-1}, \cdots, \omega^{a-1}) \notag \\
    \leftrightarrow \,\,
    R'_1&=(\omega^{a-1},\cdots, \omega^{a-1} \,|\, \omega^{a},\cdots, \omega^{a} \,|\, \omega^{a+1}, \cdots, \omega^{a+1}) \\
    \leftrightarrow \,\,
    R''_1&=(\omega^{a+1},\cdots, \omega^{a+1} \,|\, \omega^{a-1},\cdots, \omega^{a-1} \,|\, \omega^{a}, \cdots, \omega^{a}), \notag
\end{align}
where $n_1=n_2=n_3=n/3$, $m_1=m_2=m_3$ and $m'_1=m'_2=m'_3$ are satisfied.
We notice that these permutations are just a repetition of the 3-permutation (\ref{T2Z3www_normal}).


\section*{Acknowledgment}
The authors would like to thank Kenta Kojin for useful discussions. We are indebted to members of our laboratory for encouragements.

\bibliographystyle{unsrt} 
\bibliography{main} 

\end{document}